# Effects of residual stress on the isothermal tensile behavior of nanocrystalline superelastic NiTi shape memory alloy

Kai Yan[1, 2, *], Pengbo Wei[1, 2], Weifeng He[3,4], Qingping Sun[2]

[1]Department of Materials Science and Engineering, Southern University of Science and Technology, Shenzhen 518055, Guangdong, China

[2]Department of Mechanical and Aerospace Engineering, The Hong Kong University of Science and Technology, Hong Kong, China

[3]State Key Laboratory for Manufacturing Systems Engineering, School of Mechanical Engineering, Xi'an Jiaotong University, Xi'an 710049, China

[4]Science and Technology on Plasma Dynamics Laboratory, Air Force Engineering University, Xi'an 710038, China

**Abstract**

The residual stress greatly affects the mechanical behavior of a material. In this work, the effect of residual stress on the isothermal tensile behavior of a NiTi shape memory alloy is studied. The focused ion beam and digital image correlation are combined to measure the two-dimensional residual stress in nanocrystalline NiTi plates processed with prestrain laser shock peening. A four-point bending experiment verified the accuracy of this measurement method. The FIB-DIC method is an attractive tool for measuring the two-dimensional residual stress in phase transition nanocrystalline materials. The internal residual stress significantly decreases the phase transition stress, and the mechanism is studied via finite element and theoretical analyses. This work implies that the mechanical behavior of NiTi shape memory alloys can be tailored via residual stress engineering.

*Corresponding authors.

E-mail address: kyanaa@connect.ust.hk (K. Yan).

**1. Introduction**

The superelastic and shape memory effects are considered two of the most researched topics in academic research and industry because of the subtle properties of NiTi shape memory alloys (SMAs). These subtle properties provide the ability of NiTi alloys to return to their original undeformed shape upon unloading or heating, which places the materials in a privileged position for applications in fields such as biomedical devices, aerospace structures, and actuation systems. Specifically, they find application in shock absorbers, antiscald valves, bone implants, and cardiovascular stents, where precise mechanical response and recovery are essential [1-3]. In contrast, while NiTi alloys have wide applications, their usage has often been circumscribed through early failure during cyclic loading as a result of their limited tensile and bending fatigue life [4-6]. This intrinsic limitation prevents them from performing much better in applications requiring high reliability and durability for a long period of time.

The techniques used to improve the fatigue life of NiTi SMAs include the following: a few promising techniques induce compressive residual stresses on the material surface to delay fatigue crack initiation and propagation, which ultimately improve material performance under cyclic loading [7, 8]. This is supported by the fact that although laser shock peening is a well-established surface treatment technique applied in a wide manner within both the aerospace and automotive industries, it has properties of geometry-independent applicability, improved fatigue resistance, surface nanocrystallization, and generation of compressive residual stress [9-12]. The application of LSP imparts a high strain rate plastic deformation near the surface of the material, which induces the nanocrystalline layer and results in a beneficial compressive residual stress state. Such effects have been shown to be crucial for enhancing the fatigue life of most metallic systems tested, including NiTi SMAs [13].

In most cases, residual stresses are developed from various techniques, such as LSP, mechanical working, and heat treatment, that significantly affect the mechanical behavior of the material of interest. Accurate measurements of these stresses are urgently needed in nature, especially in complicated systems such as nanocrystalline materials, to understand and optimize the performance of treated components. Traditional residual stress measurement techniques can be broadly classified into specific categories: destructive, nondestructive, and semidestructive techniques. However, most of the destructive techniques utilized for such tests, such as hole-drilling [14] and ring-core techniques [15], have been applied mainly to large-scale components and do not have the precision required for thin films or small-scale regions. Many noncontact techniques include X-ray diffraction (XRD) [16], synchrotron X-ray [17], and neutron diffraction [18], but all these techniques have other problems and provide only more detailed information on residual stress states. For example, XRD is limited to surface stress measurements and is applicable only for flat workpieces. Synchrotron and neutron diffraction techniques, though available, are restricted by the unavailability and high running costs of these facilities, which makes them very expensive and hence not practical for general usage [19].

In the last decade, various techniques have been developed for the micro- and nanomeasurement of residual stresses. Focused ion beams coupled with digital image correlation have emerged as powerful tools for mapping two-dimensional residual stress distributions with high precision. As a semidestructive testing method, FIB-DIC is particularly suitable for small-scale materials and components, such as thin plates or nanocrystalline regions, where many traditional techniques cannot normally be applied. The described combination of FIB and DIC enables localized residual stress measurements with a resolution at the level of microns, providing insight into aspects not accessible with other techniques.

In this study, the effects of residual stress on the isothermal tensile properties of nanocrystalline NiTi-shaped memory alloys were investigated. FIB-DIC is used to quantitatively measure the two-dimensional residual stress distribution in NiTi SMA samples subjected to LSP at different levels of prestrain, namely, 0%, 6%, and 9%, and the effect of residual stress on the isothermal tensile behaviors of NiTi SMA is studied. Experimental techniques, finite element analysis, and theoretical models were combined to investigate the relationship between residual stress and phase transition stress. This study provides evidence that residual stress engineering has the potential to tailor the valuable mechanical behavior of NiTi SMAs, effectively influencing extensive practical applications and enhancing the reliability of these metals during service.

## 2. Materials and experiments

The polycrystalline 50.9 at.% Ni-49.1 at.% Ti sheet obtained from Johnson Matthey Medical Components, Inc., USA, used in this work had a thickness of 1.50 mm. The austenite finish temperature is 287 K, and the material is superelastic at room temperature. The microstructure of the as-received material was examined via transmission electron microscopy (TEM, Tecnai F30). The TEM sample was prepared via ion milling via a Fischione Model 1050 ion thinning instrument. Fig. 1a shows that the mean crystalline size of the as-received NiTi plate is approximately 65 nm. The selected area diffraction pattern in Fig. 1b shows that the as-received sample has the B2 phase at room temperature [20].

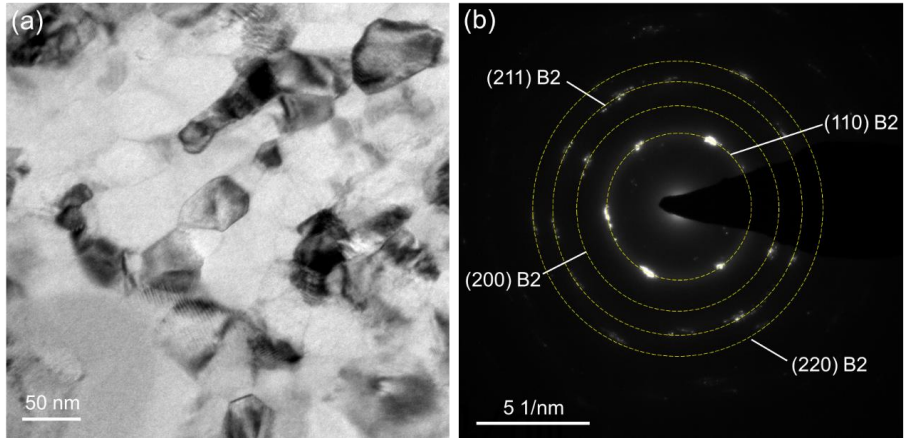

**Fig. 1.** (a) Bright-field TEM image and (b) SAED pattern of the as-received NiTi SMA.

NiTi plates with a thickness of approximately 1.42 mm after polishing were wire-cut and mechanically polished to mirror-like surfaces with abrasive paper followed by an $Al_2O_3$ suspension. Then, the dog-bone-shaped samples with gauge lengths of 20 mm were subjected to prestrain LSP. The samples were prestrained to 6% and 9% strain separately before LSP treatment, and the strain was calibrated with the DIC method. According to the experimental results of the as-received NiTi, the material fully transforms to martensite with a 6% prestrain. Thereafter, the samples were subjected to the Q-switched Nd:YAG laser under a laser intensity of 7.9 $GW/cm^2$, an overlap ratio of 56.5%, a beam diameter of 2 mm, and three shock cycles (Fig. 2). The copper foil was glued at the back of the NiTi sample to prevent possible spallation of the sample surface. Water was used as an absorbent overlay, which is transparent in nature and is not an appropriate material for use during LSP. A single-crystal NiTi strip (with a thickness of approximately 0.9 mm) was subjected to a calibration experiment. The as-received NiTi strip was annealed at 500 °C for 5 min to relax any possible residual stress in the material. A calibration experiment was also carried out with a superelastic nanocrystalline NiTi strip. A universal tensile machine equipped with an extensometer was used to measure the tensile stress-strain curves of the dog bone-shaped specimens, the loading rate was set as $1.3 \times 10^{-4}$ s$^{-1}$ with a maximum displacement of 1.3 mm.

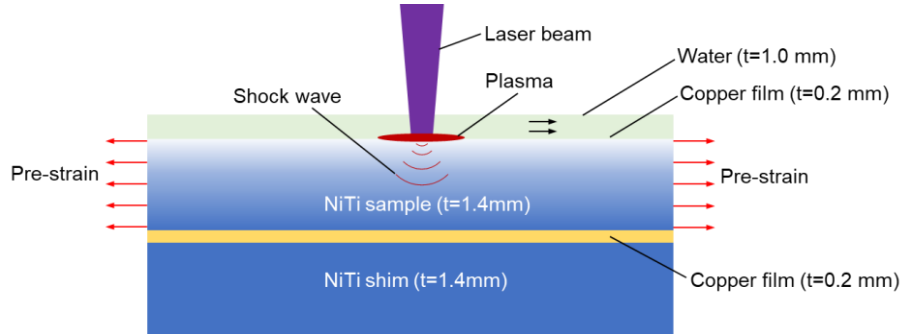

**Fig. 2.** Schematic illustration of prestrain laser shock peening processing.

The measurements were carried out in an FEI Nanolab-600i system with focused ion beam-digital image correlation (FIB-DIC) tests. A schematic of the measurement process is presented in Fig. 3a and 3b. Prior to each test, the sample was embedded in epoxy resin and mechanically polished on the side surface. Further electrolytic polishing was conducted to eliminate the possibility that any residual stress layers were introduced during mechanical polishing. A thin, ~8 nm gold coating was deposited on the prepared surface via a Scancoat Six (Edwards) sputter coater under a scanning electron microscope (SEM). The scanning resolution was 1536 × 1024 pixels, with a frame time of 180 ms, line time of 164.4 µs, and dwell time of 100 ns. Drift correction imaging was used as part of the technique for drift control imaging during analysis. An 11 × 11 grid of points (100 nm diameter, spaced at 1 µm) was milled for 200 µs using a current of 0.23 nA. In the center, a larger reference mark with a diameter of 200 nm was scribed. The first image, which was obtained at 26 pA and a magnification of 25,000 times, is shown in Fig. 3c. A 400 s milling at 9.3 nA created a rectangular trench with an inner length of 12 µm and an outer length of 13 µm. Following the milling of a trench, the same area on the sample was imaged with ion beam imaging, using the same parameters as in the first image (see

Fig. 3d). Digital image correlation (DIC) was used to identify the displacement of the reference marks: 9 points with an original spacing of approximately 8 μm were chosen for the relief strain estimation to minimize the effects of Ga+ ion damage at the trench edges [21, 22].

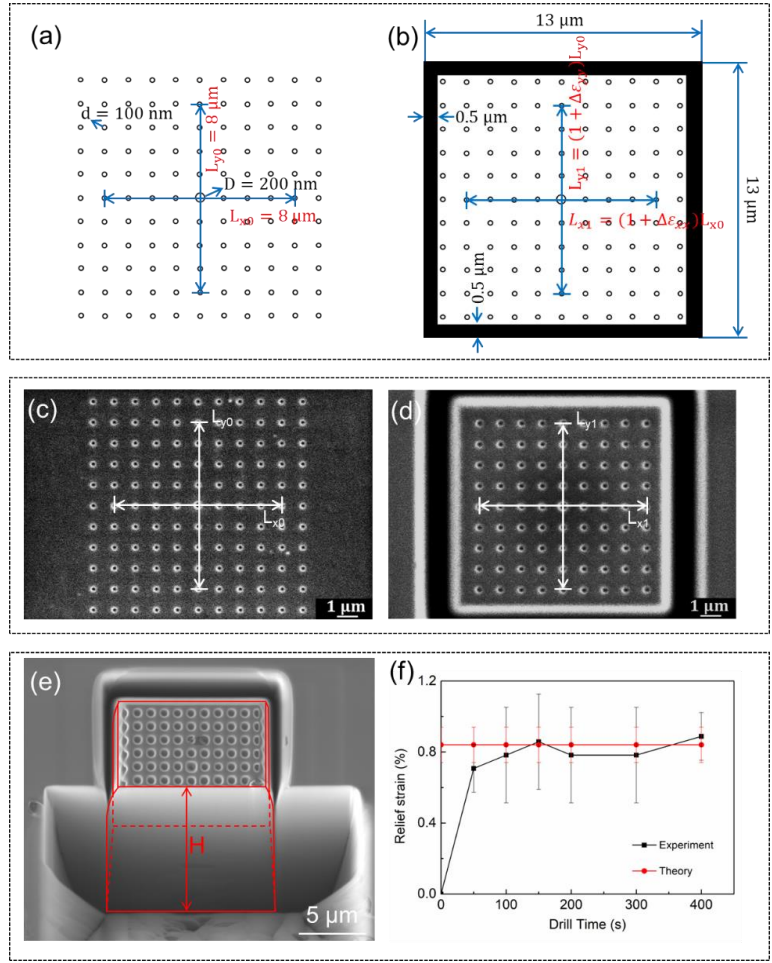

**Fig. 3.** (a) Schematic of the dot matrix pattern prior to residual stress relaxation. (b) Schematic of the dot matrix pattern after residual stress relaxation. (c) Ion image of the FIB-fabricated dot matrix on the top surface of the micropillar before residual stress relaxation by trench cutting. (d) Ion image of the FIB-fabricated dot matrix on the top surface of the micropillar after residual stress relaxation by trench cutting. (e) SEM image of the rectangular micropillar for residual stress evaluation and (f) variation in the measured residual elastic strain with drilling depth H.

## 3. Experimental results

Therefore, an *in situ* calibration experiment was conducted on an FIB system under a four-point bending loading frame to verify the feasibility of the method. The superior superelasticity performance of the single-crystal NiTi strip—without obvious residual strain after the unloading process—was clearly demonstrated by a series of four-point bending tests before the calibration process. Fig. 3e presents an SEM image of a rectangular pillar measured for strain relief. The depth of drilling H ≈ 21.5 μm. In Fig. 3f, the measured relief strain is compared with the theoretical prediction strain as a function of the drilling depth. The measured relief strain was derived in the *x*- and *y*-directions according to Equations (1) and (2).

$$\Delta \varepsilon_{xx} = \frac{\Delta L_x}{L_{x0}} = \frac{L_{x1} - L_{x0}}{L_{x0}} \qquad (1)$$

$$\Delta \varepsilon_{yy} = \frac{\Delta L_y}{L_{y0}} = \frac{L_{y1} - L_{y0}}{L_{y0}} \qquad (2)$$

where $L_{x0}$ and $L_{y0}$ represent the original distance between two marks, $\Delta \varepsilon_{xx}$ and $\Delta \varepsilon_{yy}$ represent the measured relief strain along the *x*- and *y*-directions, $L_{x1}$ and $L_{y1}$ represent the distance between two marks after drilling a rectangular trench, and $\Delta L_x$ and $\Delta L_y$ represent the distance change before and after drilling. As shown in Fig. 3f, the measured value of the relief strain increases with increasing drilling depth. It remains almost constant (same as the theoretical value) when the drilling depth is greater than 10 μm, meaning that the residual stress is fully relaxed.

A schematic of the calibration experiment is shown in Fig. 4. Fig. 4a shows the schematic setup for a four-point bending calibration experiment. The theoretical value of 0.84 ± 0.10%, which was calculated via Equation (1), as explained later, is shown in Fig. 4b. When the depth H of the drilling increases, the relief strain also increases, and above a depth of 21.5 μm, the measured values are in agreement with the theoretical prediction.

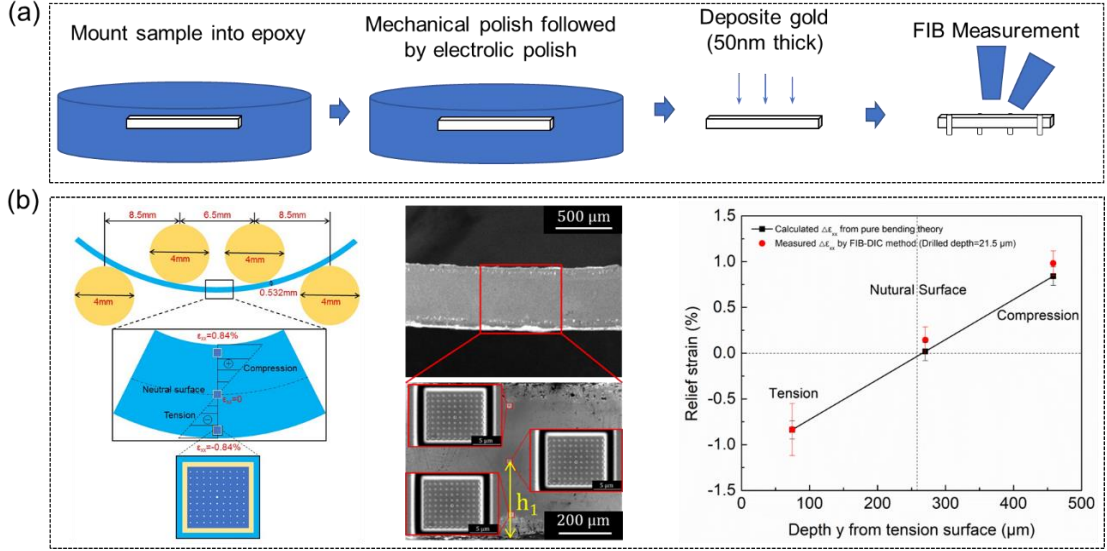

**Fig. 4.** (a) Schematic diagram of the four-point bending calibration experiment set-up, (b) schematic diagram and SEM images of the four-point bending calibration experiment with an annealed single-crystal NiTi strip, and comparison between the calculated and measured relief strains in the *x* direction.

The theory for calculating the relief strain in pure bending is shown in formula (3):

$$\Delta\varepsilon = \frac{h}{R} \qquad (3)$$

where $\Delta\varepsilon$ is the nominal relief strain and where $h$ is the distance between the measured location and the neutral surface (negative for the tensile side and positive for the compressive side). $R$ is the radius of the neutral surface, which can be calculated from the SEM image of the side surface through fitting, and the calculated value of $R$ is approximately $20.58 \pm 2.82$ mm. The maximum relief strain at a depth of 75 μm is approximately $0.83 \pm 0.28\%$ (negative for the tensile side and positive for the compressive side). The calculated relief strain at the depth of 75 μm is approximately $0.84 \pm 0.10\%$ (negative for the tensile side and positive for the compressive side).

A comparison between the calculated result with formula (3) and the measured result with the FIB-DIC method is shown in Fig. 4b. The experimental results coincide with the theoretical prediction very well, thus verifying the accuracy of this method. Similar results were also found for polycrystalline NiTi. Notably, the measured average value at the neutral surface is approximately 0.2%, which may be caused by the surface damage induced by $Ga^+$ ions [21, 22].

After calibrating the applicability of the FIB-DIC method for the NiTi material, this method was used to measure the distributions of two-dimensional relief strain and residual elastic strain in nanocrystalline NiTi after LSP processing. Fig. 5 shows a schematic diagram of the LSP treatment and FIB drilling configuration and SEM images of the side surface of LSP-treated NiTi.

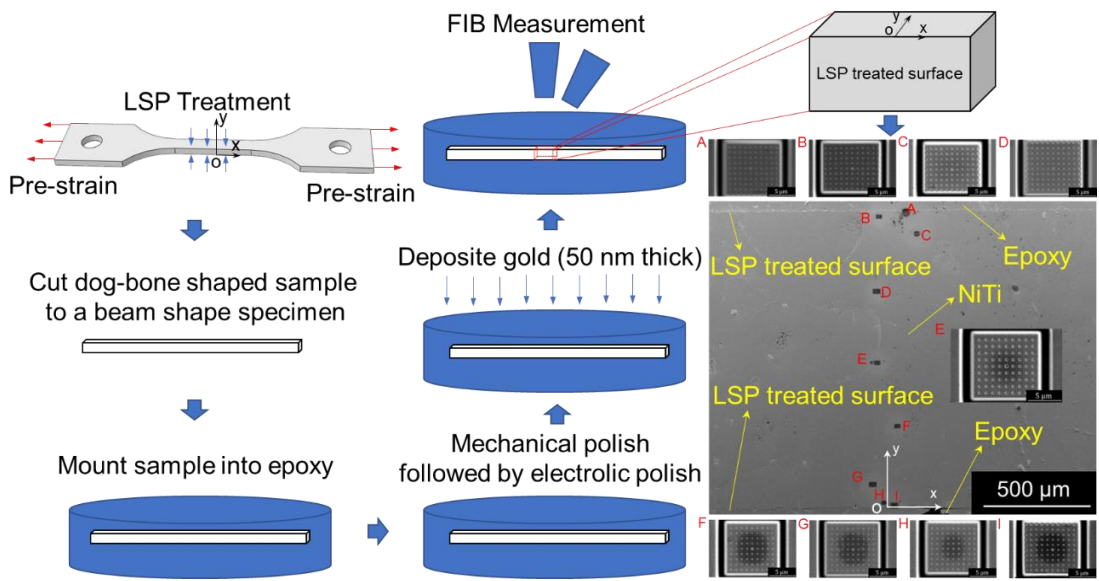

**Fig. 5.** Schematic diagram of the LSP treatment and FIB drilling configuration and SEM images of the side surface of LSP-treated NiTi.

Fig. 6 shows the measured results of $\Delta\varepsilon_{xx}$ (relief strain along the *x*-direction) and $\Delta\varepsilon_{yy}$ (relief strain along the *y*-direction) at different depths *y* from the bottom surface treated by LSP. Fig. 6a shows the as-received sample (AR), Fig. 6b shows the sample after LSP treatment with 6% prestrain (P6), and Fig. 6c shows the sample after LSP treatment with 9% prestrain (P9). For the as-received sample, both $\Delta\varepsilon_{xx}$ and $\Delta\varepsilon_{yy}$ are very small and close to zero. However, for the dog-bone sample processed with 6% prestrain, the maximum $\Delta\varepsilon_{xx}$ can reach 3.7%. The thickness of the layer with compressive residual elastic strain and residual stress is approximately 100 μm, and $\Delta\varepsilon_{xx}$ is approximately -0.2% in the middle part of the sample. For $\Delta\varepsilon_{yy}$, the minimum relief strain can reach -3.2%, and the relief strain is approximately 0.2% on the neutral surface of the sample. For the dog-bone sample processed with 9% prestrain, the maximum $\Delta\varepsilon_{xx}$ can reach 3.5%. The thickness of the layer with compressive residual elastic strain and residual stress is approximately 100 μm, and $\Delta\varepsilon_{xx}$ is approximately -0.4% in the middle part of the sample. For $\Delta\varepsilon_{yy}$, the minimum relief strain can reach -2.8%, and the relief strain is approximately zero on the neutral surface of the sample.

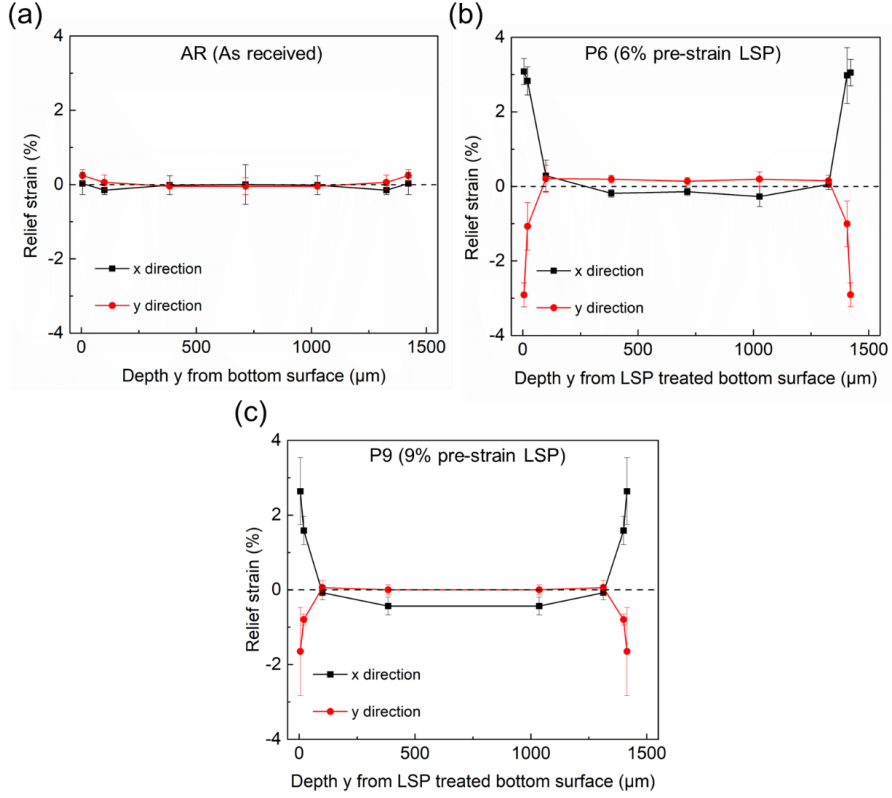

**Fig. 6.** Measured results of the relief strain in the *x*- and *y*- directions at different depths *y* from the bottom surface. (a) AR, (b) P6, and (c) P9 samples.

After $\Delta\varepsilon_{xx}$ and $\Delta\varepsilon_{yy}$ are measured, the residual stress along the *x*- and *y*-directions can be determined from formulas (4-5) [23].

$$\sigma_{res}^{xx} = -\frac{E}{(1-v^2)}(\Delta\varepsilon_{xx} + v\Delta\varepsilon_{yy}) \qquad (4)$$

$$\sigma_{res}^{yy} = -\frac{E}{(1-v^2)}(\Delta\varepsilon_{yy} + v\Delta\varepsilon_{xx}) \qquad (5)$$

where $\Delta\varepsilon_{xx}$ and $\Delta\varepsilon_{yy}$ represent the measured relief strain along the *x*- and *y*-directions, *E* is the Young's modulus of the material, $\sigma_{res}^{xx}$ and $\sigma_{res}^{yy}$ represent the residual stress along the *x*- and *y*-directions, respectively, and $v$ represents the Poisson's ratio.

From Hooke's law, the relationship between residual stress ($\sigma_{res}^{xx}$ and $\sigma_{res}^{yy}$) and residual elastic strain ($\varepsilon_{res}^{xx}$ and $\varepsilon_{res}^{yy}$) can be obtained, as shown in formulas (6-7) [24].

$$\sigma_{res}^{xx} = E\varepsilon_{res}^{xx} \qquad (6)$$

$$\sigma_{res}^{yy} = E\varepsilon_{res}^{yy} \qquad (7)$$

The relationship between the relief strain and residual elastic strain is as follows:

$$\varepsilon_{res}^{xx} = -\frac{\Delta\varepsilon_{xx}+\nu\Delta\varepsilon_{yy}}{1-\nu^2} \qquad (8)$$

$$\varepsilon_{res}^{yy} = -\frac{\Delta\varepsilon_{yy}+\nu\Delta\varepsilon_{xx}}{1-\nu^2} \qquad (9)$$

where $\varepsilon_{res}^{xx}$ and $\varepsilon_{res}^{yy}$ represent the residual elastic strains in the *x*- and *y*-directions, respectively.

Fig. 7a-c shows the calculated residual elastic strains along the *x* and *y* directions at different depths *y* from the bottom surface of the AR, P6, and P9 samples, respectively. The residual elastic strain for the AR sample is minimal. In contrast, for the P6 sample, the residual elastic strain along the *x*-direction is negative at the laser processed surface. There is a minimum value of approximately -3.4% at a depth of approximately 20 μm from the LSP processed surface, and the minimum value increases slightly (-2.9%) at the LSP treated surface. It changes to positive (~0.2%) in the middle part of the sample. This result means that along the *x*-direction, there is very high compressive residual stress at the LSP-treated surface and tensile residual stress in the middle region of the sample. However, the value of $\varepsilon_{res}^{yy}$ decreases dramatically from the LSP-treated surface (maximum 2.6%) and becomes negative (-0.1%) in the middle region of the sample. This means that along the *y*-direction, there is very high tensile residual stress at the LSP-treated surface and compressive residual stress in the middle.

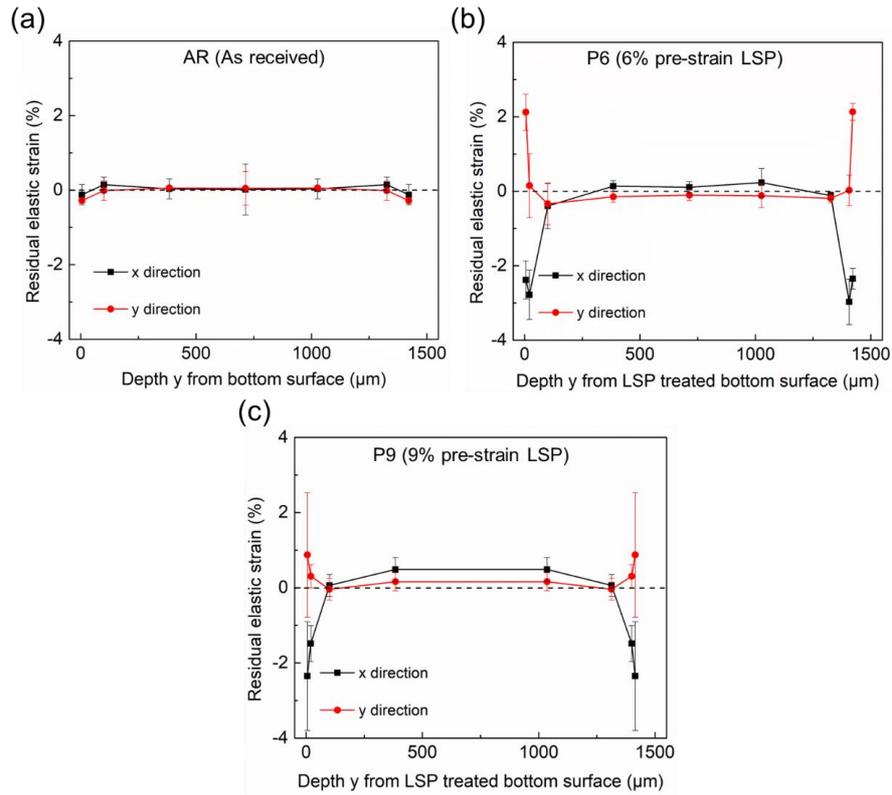

**Fig. 7.** Calculated results of residual elastic strain in the *x*- and *y*-directions at different depths y from the LSP-treated bottom surface. (a) AR, (b) P6, and (c) P9 samples.

For the P9 sample, the residual elastic strain along the *x*-direction is negative at the laser-processed surface. In comparison, there is a minimum value of approximately -3.8% at a depth of approximately 5 μm from the LSP processed surface, and the value changes to positive (~0.5%) in the middle part of the sample. This result means that along the *x*-direction, there is very high compressive residual stress at the LSP-treated surface and tensile residual stress in the middle region of the sample. However, the value of $\varepsilon_{res}^{yy}$ decreases dramatically from the LSP-treated surface (maximum 2.5%) and changes to 0.2% in the middle region of the sample. This means that along the *y*-direction, there is very high tensile residual stress across the sample.

Fig. 8 shows the measured results of the Young's modulus at different depths *y* from the bottom surface of the AR, P6, and P9 samples. The Young's modulus is approximately 60 GPa for the AR sample, whereas for the P6 sample, the value can reach

~80 GPa at the LSP-treated surface; the value decreases to ~60 GPa at a depth of 20 μm and remains constant with increasing depth. For the P9 sample, the value can reach ~73 GPa at the LSP-treated surface, and the value decreases to ~60 GPa at a depth of 20 μm and remains constant with increasing depth.

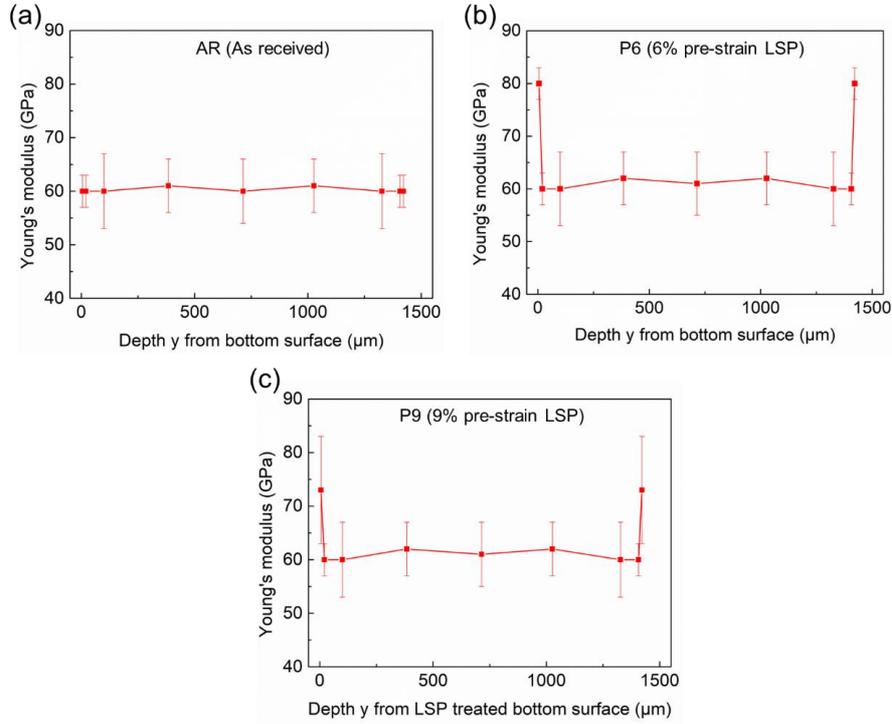

**Fig. 8.** Distribution of the Young's modulus at different depths from the bottom surface of the (a) AR, (b) P6, and (c) P9 samples.

Fig. 9 shows the calculated results of the residual stress in the *x* and *y* directions at different depths *y* from the bottom surface of the AR, P6, and P9 samples. The residual stress of the AR sample is close to zero (Fig. 9a). In contrast, for the P6 sample, the residual stress along the *x*-direction is negative at the LSP-treated surface; the minimum value is approximately -1.7 GPa at a depth of approximately 20 μm from the LSP-treated surface (Fig. 9b). It changes to positive (approximately 65 MPa) in the middle region of the sample. This means that along the *x*-direction, there is a very high compressive residual stress at the LSP-treated surface and tensile residual stress in the middle region of the sample. However, the residual stress in the *y* direction decreases dramatically from the LSP processed surface and becomes negative in the middle region of the sample. The value is positive at the LSP-treated surface, and there is a maximum value of approximately 1.6 GPa at the top surface from the LSP-treated surface. It decreases to approximately -62 MPa in the middle region of the sample.

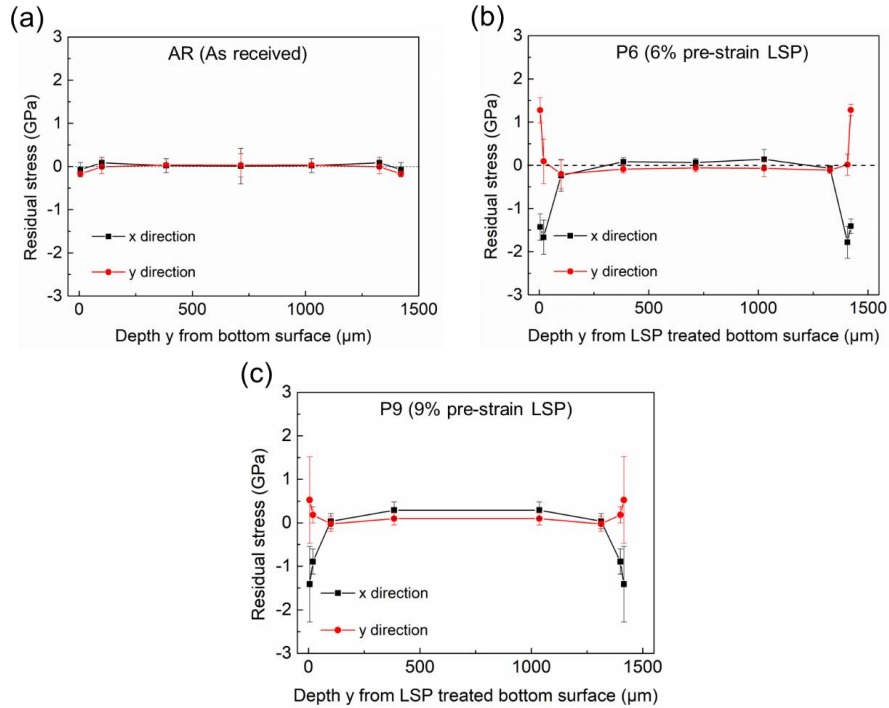

**Fig. 9.** Calculation results of residual stress in the x and y directions at different depths y from the LSP-treated bottom surface for the (a) AR, (b) P6, and (c) P9 samples.

For the P9 sample, the residual stress along the *x*-direction is negative at the LSP-treated surface; the minimum value is approximately -2.3 GPa at a depth of approximately 5 μm from the LSP-treated surface (Fig. 9c). It changes to positive (approximately 292 MPa) in the middle region of the sample. This means that along the *x*-direction, there is a very high compressive residual stress at the LSP-treated surface and tensile residual stress in the middle region of the sample. However, the value in the *y* direction decreases dramatically from the LSP processed surface to the middle region of the sample. The residual stress in the *y*-direction is positive at the LSP-treated surface, and there is a maximum value of approximately 1.5 GPa at the top surface from the LSP-treated surface. It decreases to approximately 96 MPa in the middle region of the sample.

During the LSP treatment, for the AR sample, NiTi will undergo pseudoelastic deformation, and the residual stress mainly comes from phase transition-induced plasticity, which is very low. For the P6 sample, since a prestrain of 6% is applied, austenite transforms into the martensite phase before LSP treatment, and plastic deformation of martensite more easily occurs. Since $\sigma_{HEL}$ decreases dramatically, plastic deformation of martensite occurs more easily, and the residual elastic strain accumulates with increasing number of shocks, making the measured value more significant than that in the no prestrain condition. For the P9 sample, plastic deformation of martensite is much easier than that of the P6 sample; thus, the resulting residual stress becomes more significant than that of the P6 sample.

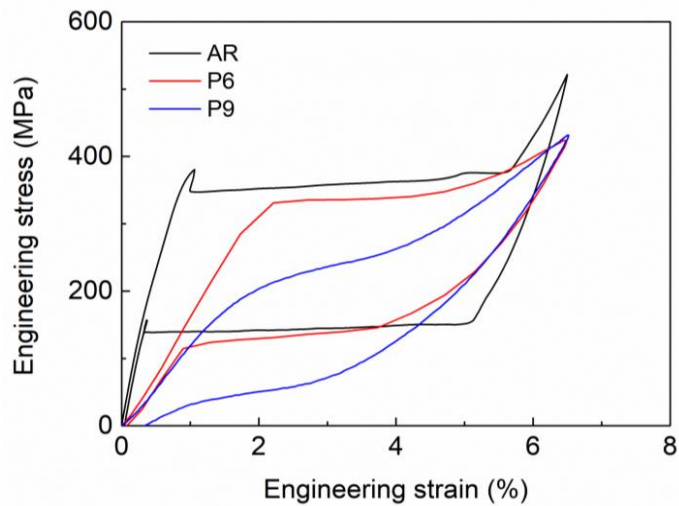

**Fig. 10.** Tensile stress–strain behavior of the AR, P6 and P9 samples under a loading strain rate of $1.3 \times 10^{-4}$ s$^{-1}$.

As shown in Fig. 10, during the tensile process of the samples, since the residual stress is negligible, the residual stress has little effect on the tensile behavior of the NiTi. For the P6 and P9 samples, the transformation stress decreases with increasing residual stress.

## 4. Thermomechanically coupled constitutive model of the NiTi SMA

The development of the finite-strain thermomechanically coupled constitutive model for SMAs is based on two successive decompositions: (i) the multiplicative decomposition of the deformation gradient into elastic, inelastic, and thermal parts and (ii) the decomposition of inelastic deformation into the components corresponding to the phase transformation and martensitic reorientation. Afterwards, a thermodynamic framework is presented together with the Helmholtz free energy density, which is used to obtain constitutive relationships in such a format that guarantees thermodynamic consistency.

4.1. Kinematics

In most of these studies, the finite strain behavior of SMAs is usually modeled analogously to the well-established finite-strain elastoplastic framework [25, 26]. The deformation gradient $F$ is multiplicatively decomposed into elastic and inelastic components, expressed as $F = F_e F_{in}$. However, in the present work, in an attempt to achieve a more complete consideration of thermomechanical coupling, an extended multiplicative decomposition of $F$ into elastic $F_e$, inelastic (transformation) $F_{in}$, and thermal $F_\theta$ parts is introduced as

$$F = F_e F_{in} F_\theta, \tag{10}$$

In this context, $F_e$ is described relative to the local unstressed intermediate configuration, whereas $F_{in}$ refers to the thermally expanded configuration, and $F_\theta$ is associated with the reference configuration. $F_\theta$ is considered to represent an isotropic thermal expansion, expressed as $F_\theta = J_\theta^{\frac{1}{3}} \mathbf{1} = (det F_\theta)^{\frac{1}{3}} \mathbf{1}$.

The right Cauchy–Green tensors for elastic and inelastic deformations are given as:

$$C_e = F_e^T F_e \quad \text{and} \quad C_{in} = F_{in}^T F_{in}. \tag{11}$$

In the small-strain Zaki–Moumni (ZM) model [27], only the inelastic strain, $\varepsilon_{in}$, which is coupled with the martensite volume fraction, z, and martensite orientation strain, $\varepsilon_{ori}$, was developed as

$$\varepsilon_{in} = z\varepsilon_{ori} \qquad (12)$$

Thus, this equation demonstrated that inelastic deformation in SMAs is caused by either the phase transformation or the reorientation of martensite. Inserting the Green–Lagrange strain $E = \frac{1}{2}(C - 1)$ into this small-strain formulation gives rise to equation (12):

$$C_{in} = zC_{ori} + (1-z)\mathbf{1}. \qquad (13)$$

where $C_{ori}$ denotes the right Cauchy–Green tensor for the martensite orientation aligned with $C_{in}$ and where additive decomposition of inelastic deformation results in a better ability of the model to capture both martensite reorientation and phase transformation simultaneously.

4.2. Thermodynamic considerations

To comply with the principle of material objectivity, we assume that the Helmholtz free energy density depends on the right Cauchy–Green deformation tensors $C_e$ and $C_{ori}$, the martensite volume fraction z, the determinant of the thermal deformation $J_q$, and the temperature θ, leading to the following expression:

$$\Psi = \Psi(C_e, C_{ori}, z, J_\theta, \theta). \qquad (14)$$

Then, the time derivative of the free energy function in equation (14) is given as:

$$\dot{\Psi} = \frac{\partial \Psi}{\partial C_e}:\dot{C}_e + \frac{\partial \Psi}{\partial C_{ori}}:\dot{C}_{ori} + \frac{\partial \Psi}{\partial z}\dot{z} + \frac{\partial \Psi}{\partial J_\theta}\dot{J}_\theta + \frac{\partial \Psi}{\partial \theta}\dot{\theta}. \qquad (15)$$

The rate forms of equations (11) and (13) lead to the following results:

$$\dot{C}_e = -\frac{2}{3}J_\theta^{-1}\dot{J}_\theta C_e + J_\theta^{-\frac{2}{3}}F_{in}^{-T}\dot{C}F_{in}^{-1} - C_e L_{in} - L_{in}^T C_e, \qquad (16)$$

$$\dot{C}_{in} = 2F_{in}^T D_{in} F_{in} = \dot{z}(C_{ori} - \mathbf{1}) + z\dot{C}_{ori}, \qquad (17)$$

where $C$ is the total Cauchy–Green tensor, $L_{in} = \dot{F}_{in} F_{in}^{-1}$ is the inelastic velocity gradient and $D_{in} = \frac{1}{2}(L_{in} + L_{in}^T)$ is the inelastic deformation rate. The first term on the right-hand side of equation (17) denotes phase transformation at a fixed orientation, whereas the second term denotes martensite reorientation at a constant martensite volume fraction.

The second law of thermodynamics, expressed in the form of the Clausius–Duhem inequality, is written as:

$$S : \frac{1}{2}\dot{C} - (\dot{\Psi} + \eta \dot{\theta}) - \frac{1}{\theta} q \cdot \nabla \theta \geq 0, \tag{18}$$

where $S$ is the second Piola–Kirchhoff stress, $\eta$ is the entropy density and $q$ is the heat flux vector.

By substituting equations (15) through (17) into equation (18) and considering the coaxiality between $\partial \Psi / \partial C_e$ and $C_e$, we obtain:

$$\left( \frac{1}{2} S - J_\theta^{-\frac{2}{3}} F_{in}^{-1} \frac{\partial \Psi}{\partial C_e} F_{in}^{-T} \right) : \dot{C} + \left[ F_{in}^{-1} \frac{\partial \Psi}{\partial C_e} F_{in}^{-T} : (C_{ori} - 1) - \frac{\partial \Psi}{\partial z} \right] \dot{z} + \left( z F_{in}^{-1} C_e \frac{\partial \Psi}{\partial C_e} F_{in}^{-T} - \frac{\partial \Psi}{\partial C_{ori}} \right) : \dot{C}_{ori} + \left( \frac{2}{3} J_\theta^{-1} \frac{\partial \Psi}{\partial C_e} : C_e - \frac{\partial \Psi}{\partial J_\theta} \right) \dot{J}_\theta - \left( \eta + \frac{\partial \Psi}{\partial \theta} \right) \dot{\theta} - \frac{1}{\theta} q \cdot \nabla \theta \geq 0. \tag{19}$$

For arbitrary thermodynamic processes, the above inequality is satisfied by the following stress–strain and heat equations:

$$S = 2 J_\theta^{-\frac{2}{3}} F_{in}^{-1} \frac{\partial \Psi}{\partial C_e} F_{in}^{-T},$$

$$\frac{\partial \Psi}{\partial C_e} : C_e = \frac{3}{2} J_\theta \frac{\partial \Psi}{\partial J_\theta}, \quad \eta = -\frac{\partial \Psi}{\partial \theta} \tag{20}$$

and by the inequalities ensuring non-negative intrinsic dissipation for arbitrary changes in the internal variables z and $\dot{C}_{ori}$:

$$\begin{cases} [M : (C_{ori} - 1) - Z] \cdot \dot{z} \geq 0 \\ (zM - X) : \dot{C}_{ori} \geq 0 \end{cases} \tag{21}$$

with

$$M = F_{in}^{-1} C_e \frac{\partial \Psi}{\partial C_e} F_{in}^{-T}, \quad Z = \frac{\partial \Psi}{\partial z}, \quad \text{and} \quad X = \frac{\partial \Psi}{\partial C_{ori}} \tag{22}$$

as well as non-negative dissipation due to heat conduction:

$$-\frac{1}{\theta}\boldsymbol{q}\cdot\nabla\theta \geq 0, \tag{23}$$

where $\boldsymbol{M}$ and $\boldsymbol{X}$ are the symmetric stress-like tensors defined with respect to the thermally expanded configuration and $Z$ is a scalar variable.

The non-negative dissipation during phase transformation, as represented by equation $(21)_1$, is maintained by the following evolution equation with respect to z:

$$\dot{z} = \dot{\gamma}^z \frac{\mathcal{A}}{|\mathcal{A}|}, \tag{24}$$

where $\dot{\gamma}^z$ is a non-negative multiplier and where $\mathcal{A} = \boldsymbol{M}:(\boldsymbol{C}_{ori}-\boldsymbol{1})-Z$ represents the thermodynamic force driving the phase transformation. The associated loading function for the phase transformation is as follows:

$$\mathcal{F}_z = |\mathcal{A}| - Y_z, \tag{25}$$

where $Y_z$ controls the phase transformation threshold. $\dot{\gamma}^z$ and $\mathcal{F}_z$ are subjected to the Kuhn–Tucker consistency conditions:

$$\dot{\gamma}^z \geq 0, \quad \mathcal{F}_z \leq 0, \quad \dot{\gamma}^z \mathcal{F}_z = 0 \tag{26}$$

The dissipation due to martensite reorientation $(21)_2$ is rewritten, using a push-forward operation $\dot{\boldsymbol{C}}_{ori} = 2\boldsymbol{F}_{ori}^T \boldsymbol{D}_{ori} \boldsymbol{F}_{ori}$, in the unstressed intermediate configuration as

$$2\boldsymbol{F}_{ori}(z\boldsymbol{M}-\boldsymbol{X})\boldsymbol{F}_{ori}^T : \boldsymbol{D}_{ori} \geq 0 \tag{27}$$

Following the principle of maximum dissipation in elastoplasticity [28], the associative evolution equation for martensite reorientation is as follows:

$$\boldsymbol{D}_{ori} = \frac{1}{2}\dot{\gamma}^t \frac{\boldsymbol{\mathcal{B}}^D}{||\boldsymbol{\mathcal{B}}^D||} \geq 0 \tag{28}$$

where $\dot{\gamma}^t$ is a non-negative multiplier, $(\cdot)^D = (\cdot) - \frac{1}{3}tr(\cdot)$ is used to extract the deviator of a tensor, and $\boldsymbol{\mathcal{B}} = \boldsymbol{F}_{ori}(z\boldsymbol{M}-\boldsymbol{X})\boldsymbol{F}_{ori}^T$ denotes the thermodynamic force associated with martensite reorientation. To express the constitutive equations in a Lagrangian frame, the evolution equation (28) is mapped into the thermally expanded configuration through a pull-back operation, resulting in:

$$\dot{\boldsymbol{C}}_{ori} = \dot{\gamma}^t \frac{\mathcal{C}^D}{||\mathcal{C}^D||} \boldsymbol{C}_{ori}, \qquad (29)$$

where $\mathcal{C} = \boldsymbol{C}_{ori}(z\boldsymbol{M} - \boldsymbol{X})$ is the thermodynamic force. The corresponding loading function for martensite reorientation is given by

$$\mathcal{F}_{ori} = ||\mathcal{C}^D|| - Y_{ori}, \qquad (30)$$

where the model parameter $Y_{ori}$ defines the martensite reorientation threshold. $\dot{\gamma}^t$ and $\mathcal{F}_{ori}$ are subjected to the Kuhn–Tucker conditions:

$$\dot{\gamma}^t \geq 0, \qquad \mathcal{F}_{ori} \leq 0 \quad , \quad \dot{\gamma}^t \mathcal{F}_{ori} = 0 \qquad (31)$$

4.3. Heat equation

The local energy balance (first law of thermodynamics) is written as

$$\dot{e} - \frac{1}{2}\boldsymbol{S}:\dot{\boldsymbol{C}} - h + \nabla \cdot \boldsymbol{q} = 0, \qquad (32)$$

where e is the internal energy density, h denotes the heat source per unit volume and q is the heat flux vector per unit area in the thermally expanded configuration. The Helmholtz free energy density $\Psi$ and its time derivative $\dot{\Psi}$ are defined correspondingly as

$$\Psi = e - \eta\theta \quad \text{and} \quad \dot{\Psi} = \dot{e} - \dot{\eta}\theta - \eta\dot{\theta}. \qquad (33)$$

Then, combining equations (15), (20), (32) and (33) gives the following entropy relation:

$$\dot{\eta}\theta = h - \nabla \cdot \boldsymbol{q} + (\boldsymbol{M}:(\boldsymbol{C}_{ori} - \boldsymbol{1}) - Z)\dot{z} + (z\boldsymbol{M} - \boldsymbol{X}):\dot{\boldsymbol{C}}_{ori}. \qquad (34)$$

The internal energy and the entropy densities are assumed to depend on the same state variables as the Helmholtz free energy function i.e.,

$$e = e(\boldsymbol{C}_e, \boldsymbol{C}_{ori}, z, \boldsymbol{J}_\theta, \theta) \quad \text{and} \quad \eta = \eta(\boldsymbol{C}_e, \boldsymbol{C}_{ori}, z, \boldsymbol{J}_\theta, \theta) \qquad (35)$$

The specific heat, via equations (20)$_3$ and (33), is expressed as

$$c \stackrel{\text{def}}{=} \frac{\partial e(\boldsymbol{C}_e, \boldsymbol{C}_{ori}, z, \boldsymbol{J}_\theta, \theta)}{\partial \theta} = -\theta \frac{\partial^2 \Psi(\boldsymbol{C}_e, \boldsymbol{C}_{ori}, z, \boldsymbol{J}_\theta, \theta))}{\partial \theta^2}. \qquad (36)$$

Hence, using equations (35), (20)$_3$ and (36), one obtains

$$\theta\dot{\eta} = -\theta\left(\frac{\partial^2\Psi}{\partial\theta\partial C_e}:\dot{C}_e + \frac{\partial^2\Psi}{\partial\theta\partial C_{ori}}:\dot{C}_{ori} + \frac{\partial^2\Psi}{\partial\theta\partial z}:\dot{z} + \frac{\partial^2\Psi}{\partial\theta\partial J_\theta}:\dot{J}_\theta\right) + c\dot{\theta}. \tag{37}$$

Finally, combining equations (37) and (34) gives the following temperature evolution:

$$c\dot{\theta} = h - \nabla\cdot\boldsymbol{q} +$$

$$\underbrace{(\boldsymbol{M}:(\boldsymbol{C}_{ori}-\boldsymbol{1})-Z)\dot{z}+(z\boldsymbol{M}-\boldsymbol{X}):\dot{\boldsymbol{C}}_{ori}}_{\mathcal{D}=intrinsic\ dissipation}+\underbrace{\theta\left(\frac{\partial^2\Psi}{\partial\theta\partial C_e}:\dot{C}_e+\frac{\partial^2\Psi}{\partial\theta\partial J_\theta}\dot{J}_\theta\right)}_{\mathcal{X}=thermo-elastic\ coupling}+$$

$$\underbrace{\theta\left(\frac{\partial^2\Psi}{\partial\theta\partial C_{ori}}:\dot{C}_{ori}+\frac{\partial^2\Psi}{\partial\theta\partial z}\dot{z}\right)}_{\mathcal{Z}=thermo-transformation\ coupling}. \tag{38}$$

From this level of theoretical development and experimental analysis, the thermo-elastic coupling terms in equation (38) responsible for temperature changes due to variations in the state variables $C_e$ and $J_\theta$ were not yet well characterized [29, 30]. In addition, for the quasistatic or static loading conditions, the entropy changes in equation (37) because the irreversible variables $C_e$ and $J_\theta$ approach zero [31]. The thermoelastic coupling can be approximated to be negligible, and a factor $\omega = (1+\mathcal{X}/\mathcal{D})$ is adopted, which gives the ratio of the intrinsic dissipation that is transformed into heat. Equation (38) therefore reduces to:

$$c\dot{\theta} = h - \nabla\cdot\boldsymbol{q} + \omega\mathcal{D} + \mathcal{Z}, \tag{39}$$

where the thermotransformation coupling term $\mathcal{Z}$ reflects heat generation due to latent heat during phase transformation. The heat flux q is governed by Fourier's law:

$$\boldsymbol{q} = -k\nabla\theta, \tag{40}$$

where k is the nonnegative thermal conductivity.

4.4. Constitutive equations

The Helmholtz free energy function $\Psi$ in equation (14) is split into the following components:

$$\Psi = \Psi^e + \Psi^\theta + \Psi^{int} + \Psi^{cst}, \tag{41}$$

where $\Psi^e$ represents the hyperelastic strain energy, $\Psi^\theta$ represents the thermal contribution to the free energy [32, 33], $\Psi^{int}$ represents the interaction energy between the austenite and martensite phases [25], and $\Psi^{cst}$ represents the potential energy due to internal physical constraints.

The hyperelastic energy $\Psi^e$ is assumed to be an isotropic function of $C_e$, which permits it to be written in terms of the invariants of $C_e$ as:

$$\Psi^e = \frac{\mu}{2}\left(I_1^{C_e}\right) - \mu ln J + \frac{\lambda}{2}(ln J)^2, \tag{42}$$

where $\mu = (\frac{1-z}{\mu_A} + \frac{z}{\mu_M})^{-1}$ and $\lambda = (\frac{1-z}{\lambda_A} + \frac{z}{\mu_M})^{-1}$ are the equivalent Lamé constants depending on z in the austenite (martensite composite), $I_1^{C_e}$ is the first invariant of $C_e$, and J = det $F$ is the determinant of the deformation gradient. The dependence of $\Psi^e$ on the martensite volume fraction z enables us to define an equivalent elastic stiffness of the SMA that is dependent on the actual phase composition. In many SMA models, this is neglected [25, 34], where austenite and martensite possess identical elastic properties.

It represents the thermal energy $\Psi^\theta$, also called chemical energy [35], associated with the variation in internal energy and entropy. It is defined as:

$$\Psi^\theta = e_0 - \theta\eta_0 + zC(\theta) + c\left[\theta - \theta_0 - \theta \ln\left(\frac{\theta}{\theta_0}\right)\right], \tag{43}$$

where $e_0$ and $h_0$ are the reference internal energy and entropy at the reference temperature $\theta_0$, respectively, and $c$ denotes the isobaric heat capacity. To account for the coupling between the temperature and the phase reconfiguration, a temperature-dependent heat density may be introduced following $C(\theta) = \xi(\theta - \theta_0) + \kappa$ with $\xi$ and $\kappa$ model parameters.

The interaction energy $\Psi^{int}$ is usually determined from micromechanical or crystallographic considerations and very often expressed as a function of the martensite phase fraction and local martensitic orientation strain [27, 35]. Herein, $\Psi^{int}$ is considered to be an isotropic function of $\boldsymbol{C}_{ori}$; thus, the interaction energy is formulated in terms of z and the invariants of $\boldsymbol{C}_{ori}$ as:

$$\Psi^{int} = G\frac{z^2}{2} + \frac{z}{2}[\alpha z + \beta(1-z)]\left(I_1^{C_{ori}} - 3\right), \quad (44)$$

where $I_1^{C_{ori}}$ is the first invariant of $\boldsymbol{C}_{ori}$ and where $G$, $\alpha$ and $\beta$ are model parameters. One notices that the dependence of the interaction energy on z and the orientation strain $\boldsymbol{C}_{ori}$ enables the model to effectively capture the inelastic behavior driven by phase transformation or martensite reorientation.

By imposing physical bounds on the internal state variables z and $\boldsymbol{C}_{ori}$, the Lagrangian potential below is considered the contribution of constraint $\Psi^{cst}$ to the free energy in equation (32):

$$\Psi^{cst} = -\zeta^t(L - ||\boldsymbol{C}_{ori}||) - \zeta^f(1-z) - \zeta^r z, \quad (45)$$

where $L$ is the fully oriented strain magnitude of martensite variants, $\zeta^t$ is the Lagrange multiplier related to the unilateral constraint $||\boldsymbol{C}_{ori}|| \leq L$, and $\zeta^f$ and $\zeta^r$ are Lagrange multipliers related to the bilateral constraints $0 \leq z \leq 1$.

In conclusion, the Helmholtz free energy function $\Psi$ is given by

$$\Psi = \frac{\mu}{2}\left(I_1^{C_e} - 3\right) - \mu \ln J + \frac{\lambda}{2}(\ln J)^2 + e_0 - \theta\eta_0$$

$$+ zC(\theta) + c\left[\theta - \theta_0 - \theta \ln\left(\frac{\theta}{\theta_0}\right)\right]$$

$$+ G\frac{z^2}{2} + \frac{z}{2}[\alpha z + \beta(1-z)]\left(I_1^{C_{ori}} - 3\right)$$

$$- \varsigma^t(L - ||\boldsymbol{C}_{ori}||) - \varsigma^t(1-z) - \varsigma^t z. \quad (46)$$

The constitutive equations derived from the free energy function (46) are summarized in Table 1.

**Table 1.** Model summary.

Constitutive equations:

$$\begin{cases} \boldsymbol{S} = \mu \boldsymbol{C}_{in}^{-1} + (\lambda lnJ - \mu)\boldsymbol{C}^{-1} \\ \boldsymbol{C}_{in} = z\boldsymbol{C}_{ori} + (1-z)\boldsymbol{1} \\ \eta = \eta_0 + cln\left(\frac{\theta}{\theta_0}\right) - \xi z \\ \boldsymbol{M} = \frac{\mu}{2}\boldsymbol{C}_{in}^{-1}\boldsymbol{C}\boldsymbol{C}_{in}^{-1} + \frac{1}{2}(\lambda lnJ - \mu)\boldsymbol{C}_{in}^{-1} \\ Z = Gz + \left[(\alpha - \beta)z + \frac{\beta}{2}\right](I_1^{C_{ori}} - 3) + C(\theta) + \zeta^f - \zeta^r \\ \boldsymbol{X} = \frac{z}{2}[\alpha z + \beta(1-z)]\boldsymbol{1} + \zeta \frac{\boldsymbol{C}_{ori}}{||\boldsymbol{C}_{ori}||} \\ \mathcal{A} = \boldsymbol{M}:(\boldsymbol{C}_{ori} - \boldsymbol{1}) - Z \\ \boldsymbol{\mathcal{C}} = \boldsymbol{C}_{ori}(z\boldsymbol{M} - \boldsymbol{X}) \end{cases}$$

Evolution equations:

$$\dot{z} = \dot{\gamma}^z \frac{\mathcal{A}}{|\mathcal{A}|}, \dot{\boldsymbol{C}}_{ori} = \dot{\gamma}^t \frac{\boldsymbol{\mathcal{C}}^D}{||\boldsymbol{\mathcal{C}}^D||}\boldsymbol{C}_{ori}$$

Yield functions:

$$\mathcal{F}_z = |\mathcal{A}| - [a(1-z) + bz], \mathcal{F}_{ori} = ||\boldsymbol{\mathcal{C}}^D|| - z^2 Y$$

Kuhn–Tucker conditions:

$$\begin{cases} \dot{\gamma}^z \geq 0, \ \mathcal{F}_z \leq 0, \dot{\gamma}^z \mathcal{F}_z = 0 \\ \dot{\gamma}^t \geq 0, \ \mathcal{F}_{ori} \leq 0, \dot{\gamma}^t \mathcal{F}_{ori} = 0 \end{cases}$$

Temperature evolution:

$$c\dot{\theta} = h - \nabla \cdot \boldsymbol{q} + \omega\left[(\boldsymbol{M}:(\boldsymbol{C}_{ori} - \boldsymbol{1}) - Z)\dot{z} + (z\boldsymbol{M} - \boldsymbol{X}):\dot{\boldsymbol{C}}_{ori}\right] + \theta\xi\dot{z}$$

Fourier's law:

$$\boldsymbol{q} = -k\nabla\theta$$

In addition, the physical meanings of the model parameters used in the present model are as follows:

- a and b control the width of the hysteresis loop for z = 0 and z = 1, respectively.
- α and β determine the fractional rate of change in the orientation strain, $C_{ori}$, with stresses of z = 0 and z = 1, respectively.
- G characterizes an interaction between martensitic plates and influences the slope of

the stress–strain curve in the course of the phase transformation.

- Y is the stress level where the reorientation of martensite is initiated.

- L defines the maximum strain achieved by fully oriented martensite variants.

- $\xi$ is the slope of the phase diagram boundaries for forward and reverse transformations, which provides a measure of the variation in yield stress transformation with temperature.

- $\kappa$ is the density of heat at the completion temperature of the reversed-phase transformation.

## 5. Numerical simulation

To analyze the mechanism of the decrease in phase transformation stress under the effect of residual stress in detail, numerical simulations were conducted via Abaqus/Standard. A 2D stress plane model with a size of 20×1.4 mm was built for simplification. The overall mesh size was set as 0.05 mm, and the mesh was refined to 0.01 mm at the top 100 μm depth.

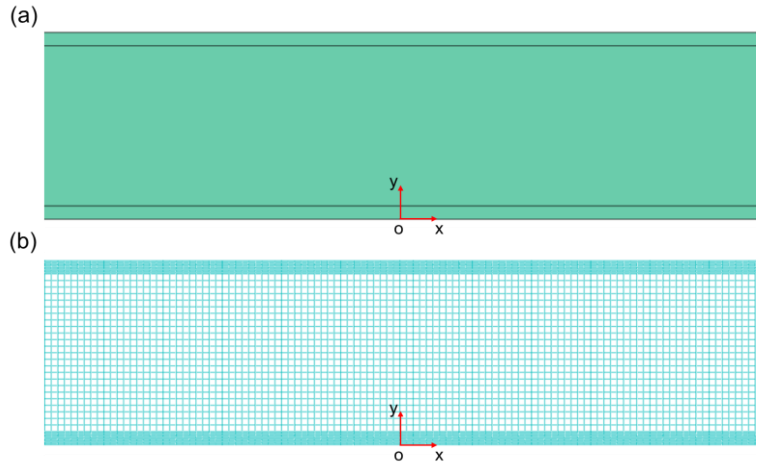

**Fig. 11.** (a) Geometry model, and (b) mesh model of the 2D NiTi specimen.

The left side of the sample was fixed, and the right side of the sample was loaded and unloaded in displacement-controlled mode. The maximum displacement was 1.3 mm, and the maximum tensile strain was 6.5%. The input residual stresses of P6 and P9 samples along the x- and y-directions are shown in Fig. 12.

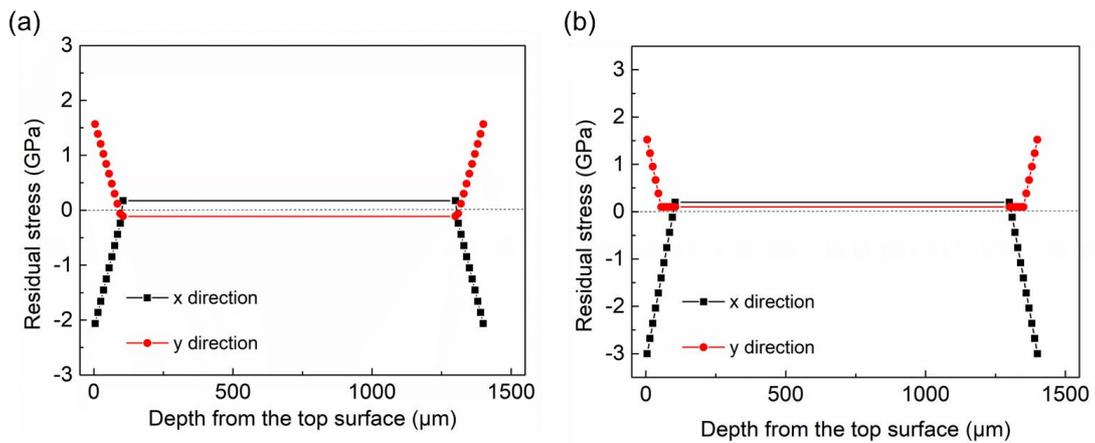

**Fig. 12.** The input residual stresses of the P6 and P9 samples along the x- and y-directions.

The simulated isothermal engineering stress–engineering strain curves of the AR, P6, and P9 samples are shown in Fig. 13. The phase transformation stress tends to decrease with increasing prestrain.

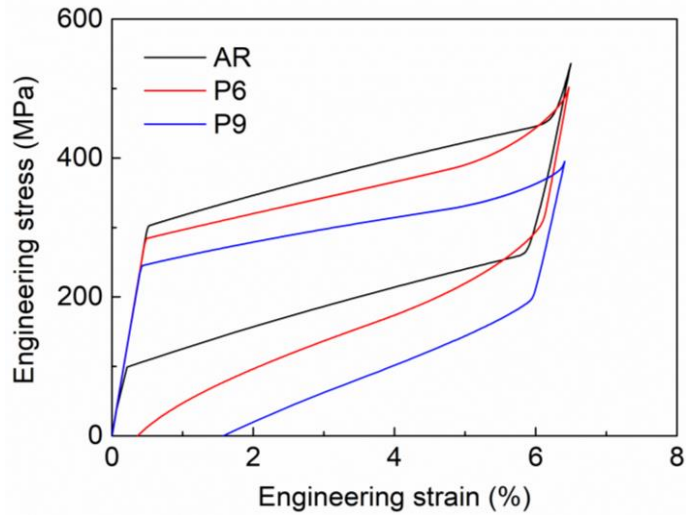

**Fig. 13.** Simulated tensile engineering stress–engineering strain behavior of the AR, P6 and P9 samples under a loading rate of 0.001 Hz.

To verify the accuracy of the simulation results, comparisons were made between the experimental and simulation results, as shown in Fig. 14. The simulation results clearly have a trend similar to that of the simulation results. The small difference may be due to many factors, including the mesh size, model constitution, and experimental error.

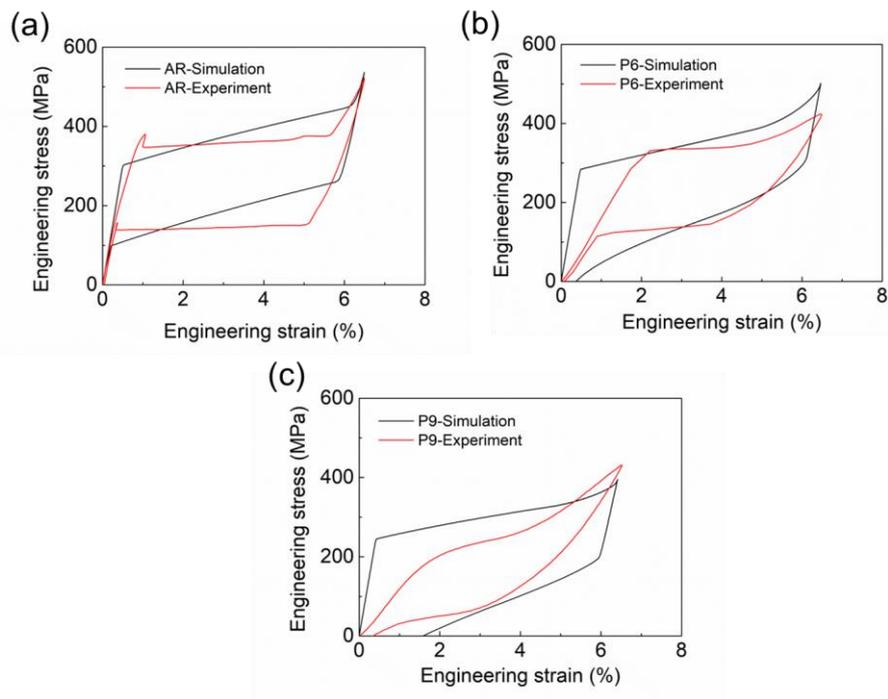

**Fig. 14.** Comparison between the simulation and experimental data. (a) AR sample. (b) P6 sample. (c) P9 sample.

To study the mechanism of the residual stress-induced decrease in the transformation stress in detail, it is essential to study the internal stress and strain fields of the samples during loading. Fig. 15 shows the contour and distribution of the von Mises stress of the cross-sections of the AR, P6, and P9 samples. The values of the von Mises stress are all positive. The stress distribution is uniform for the AR sample, and the value is approximately 535 MPa for all depths. For the P6 sample, the von Mises stress decreases from 1.79 GPa at the top surface to 148 MPa at a depth of 77 μm and increases to 714 MPa in the middle region. However, for the P9 sample, the von Mises stress decreases from 2.80 GPa at the top surface to 176 MPa at a depth of 87 μm and then increases to 624 MPa in the middle region.

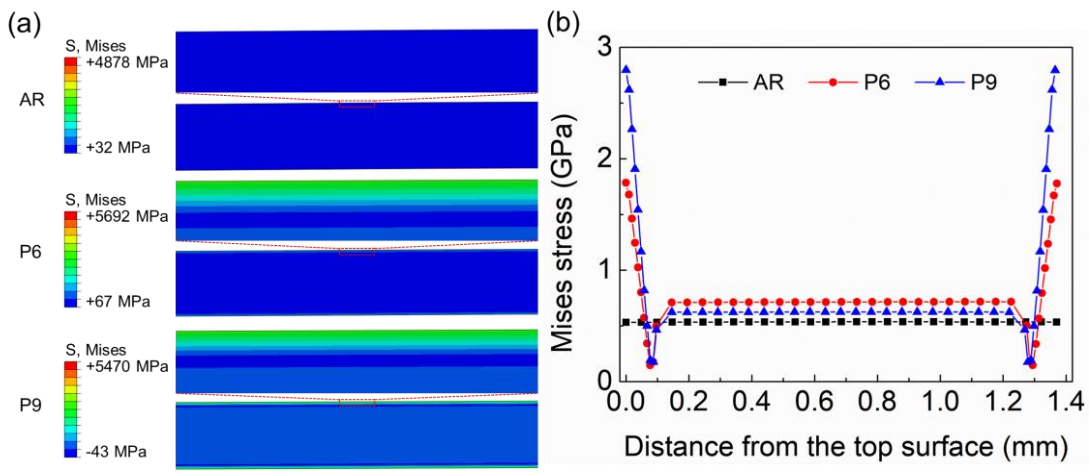

**Fig. 15.** (a) Contour of the von Mises stress field and (b) distribution of von Mises stress at different distances from the top surface of the AR, P6, and P9 samples at the maximum tensile strain.

Fig. 16 shows the contour and distribution of S11 in the cross-sections of the AR, P6, and P9 samples. The stress distribution is uniform for the AR sample, and the value is approximately 536 MPa for all depths. For the P6 sample, S11 increases from -1.26 GPa at the top surface to 680 MPa at a depth of 145 μm and remains constant in the middle region. However, for the P9 sample, S11 increases from -2.34 GPa at the top surface to 647 MPa at a depth of 145 μm and remains constant in the middle region.

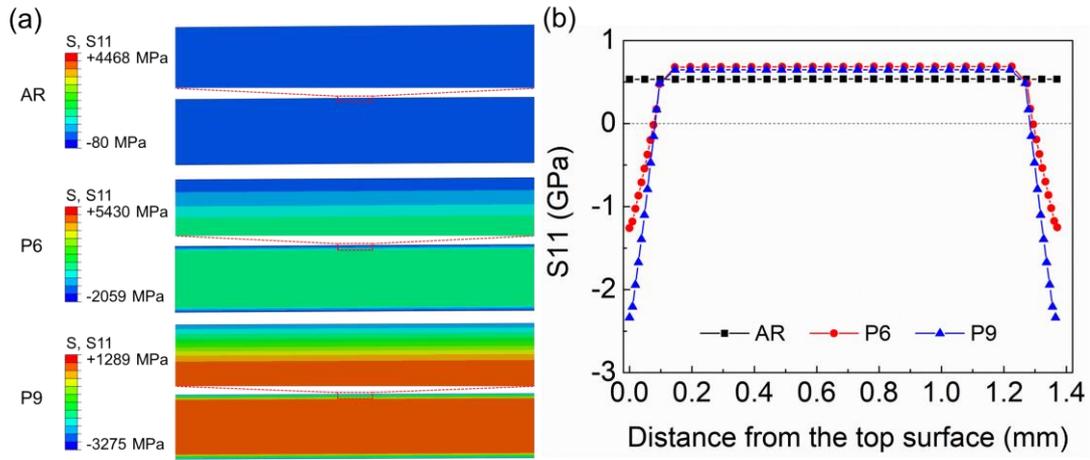

**Fig. 16.** (a) Contour of the S11 stress field and (b) distribution of S11 at different distances from the top surface of the AR, P6, and P9 samples at the maximum tensile strain.

Fig. 17 shows the contour and distribution of S22 in the cross-sections of the AR, P6, and P9 samples. The stress distribution is uniform for the AR sample, and the value is approximately zero for all depths. For the P6 sample, S22 decreases from 784 MPa at the top surface to -56 MPa at a depth of 145 μm and remains constant in the middle region. However, for the P9 sample, S11 increases from 760 MPa at the top surface to 50 MPa at a depth of 58 μm and remains constant in the middle region.

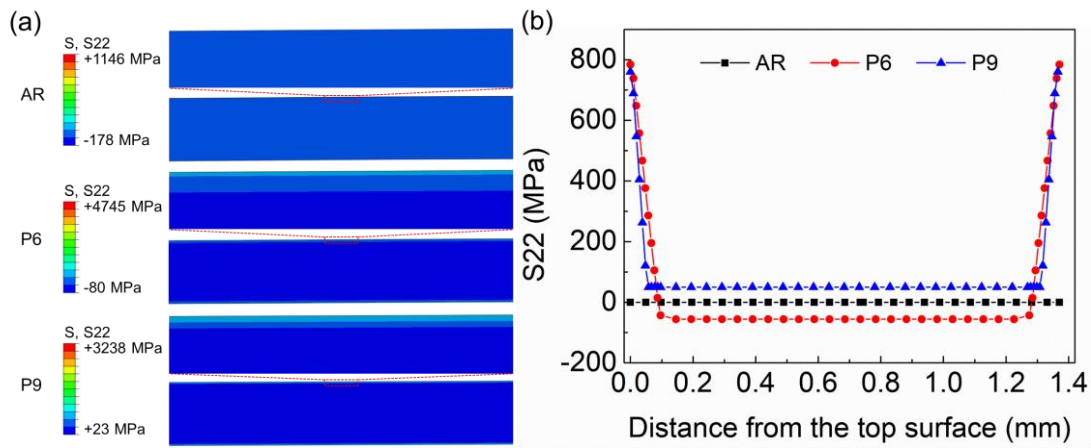

**Fig. 17.** (a) Contour of the S22 stress field and (b) distribution of true stress at different

distances from the top surface of the AR, P6, and P9 samples at the maximum tensile strain.

Fig. 18 shows the contour and distribution of LE11 in the cross-sections of the AR, P6, and P9 samples. The strain distribution is uniform for the AR sample, and the value is approximately 6.57% for all depths. For the P6 sample, LE11 increases from 6.42% at the top surface to 6.44% at a depth of 97 μm and remains constant in the middle region. However, for the P9 sample, LE11 increases from 6.38% at the top surface to 6.40% at a depth of 96 μm and remains constant in the middle region.

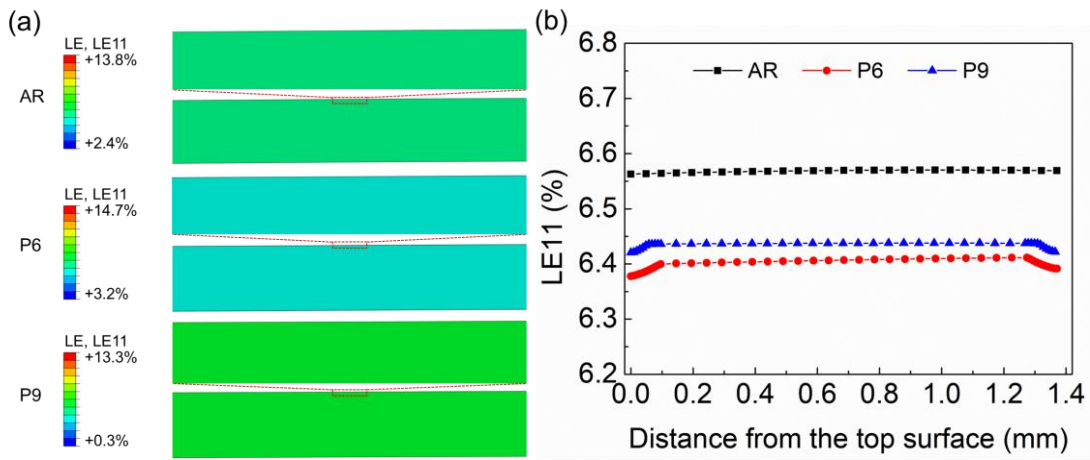

**Fig. 18.** (a) Contour of the LE11 stress field and (b) distribution of true stress at different distances from the top surface of the AR, P6, and P9 samples at the maximum tensile strain.

Fig. 19 shows the contour and distribution of LE22 in the cross-sections of the AR, P6, and P9 samples. The strain distribution is uniform for the AR sample, and the value is approximately -2.17% for all depths. For the P6 sample, LE22 increases from -5.51% at the top surface to -1.86% at a depth of 145 μm and remains constant in the middle region. However, for the P9 sample, LE22 increases from -5.73% at the top surface to -2.40% at a depth of 58 μm and remains constant in the middle region.

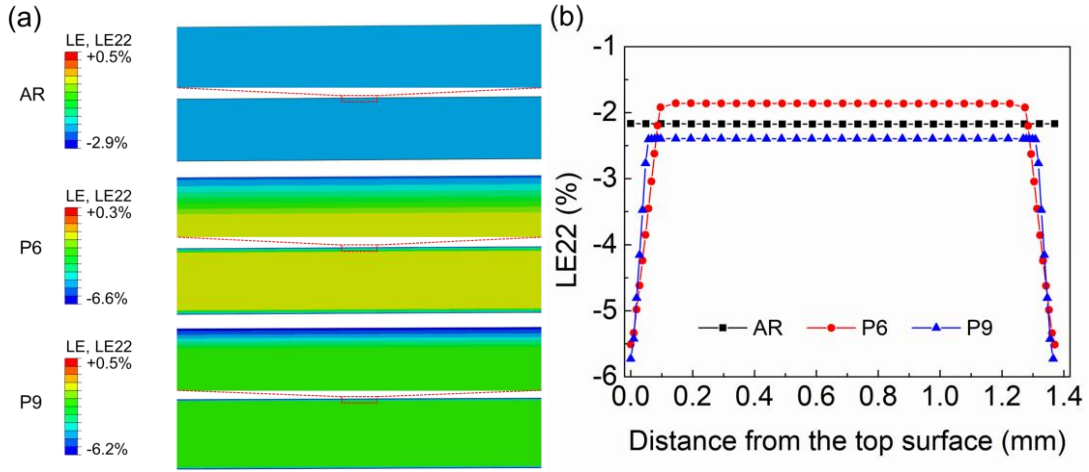

**Fig. 19.** (a) Contour of the LE22 stress field and (b) distribution of true stress at different distances from the top surface of the AR, P6, and P9 samples at the maximum tensile strain.

## 6. Discussion

The numerical simulations in this work provide significant insight into the mechanisms by which residual stress affects the phase transformation behavior of nanocrystalline NiTi SMAs under tensile loading. The current Abaqus/Standard utilizes a 2-D plane stress model to explore the effects that prestrain (induced by LSP) may have on phase transformation stresses for three conditions: AR, P6, and P9.

6.1. Influence of residual stress on phase transformation stress

All the simulation results clearly indicate that with increasing prestrain level from AR to P6 and P9, the phase transformation stress decreases continuously, as shown in Fig. 13. It then follows that fairly consistent experimental data are obtained such that residual stresses must be responsible for such a reduction in phase transformation stresses. That is, the residual stress caused by LSP changes the state of internal stresses within materials, which leads to a decrease in its phase transformation stress.

The comparison of the simulated and experimental stress–strain curves in Fig. 14 shows agreement and, therefore, validates this numerical model. Some small discrepancies in the simulation–experiment comparisons have emerged, but these discrepancies are due to limitations regarding mesh size or some constitutive material models and experimental uncertainties. In addition to these minor deviations in the simulated and experimental determinations, the overall trend remains consistent, reinforcing the accuracy of the numerical predictions.

A schematic of the mechanism of the phase transition stress decrease phenomenon of the specimen under the effect of residual stress is shown in Fig. 20. When an external stress is applied to the sample, since there is a residual tensile stress in the middle region, and there is stress concentration at the austenite/martensite interface, the stress can reach the transformation easier at the austenite/martensite interface region, triggering propagation of the martensite band, which induces nonlinear behavior in the stress–strain curve.

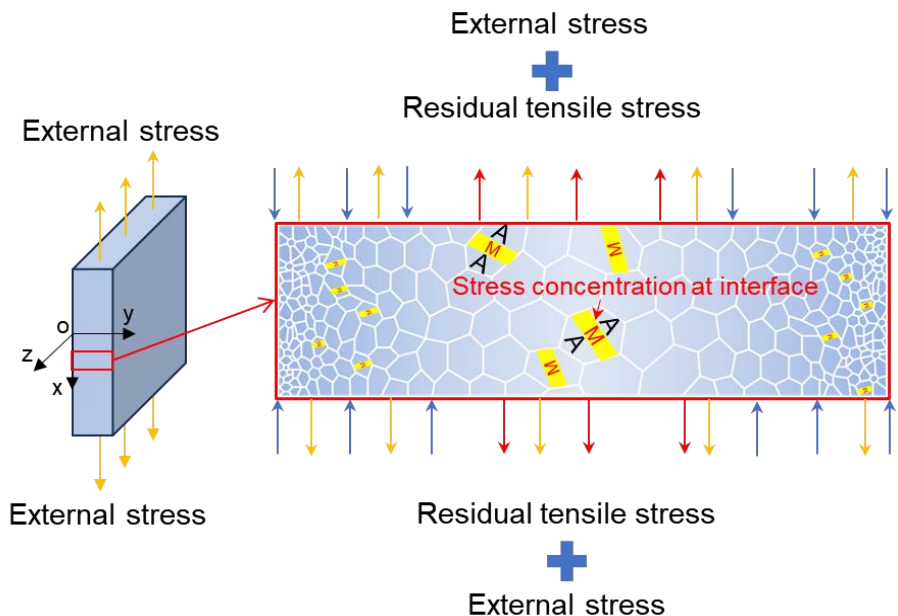

**Fig. 20.** Schematic of the mechanism of the phase transition stress decrease phenomenon of the P9 sample under the effect of residual stress.

6.2. Analysis of the stress and strain distributions

The internal stress and strain fields below further elucidate the consequences of the residual-stress state on the phase-transformation behavior. In contrast to those of the AR sample, the contour plots and distribution profiles of the von Mises stresses shown in Fig. 15 clearly illustrate that both the P6 and P9 samples exhibit a considerably nonuniform residual stress distribution. As an example, the von Mises stress values for P6 and P9 are much greater on the top surface, being 1.79 GPa and 2.80 GPa, respectively, and decrease to a moderate depth. Such a change in the profile of the internal stress distribution is most likely a direct result of residual stresses from LSP on the surface as well as in internal regions.

The S11 stress component has a very similar behavior, as shown in Fig. 16, since the AR sample is uniformly distributed, whereas P6 and P9 are nonuniformly distributed, with higher compressive stresses toward the surface. More precisely, concerning the P9 sample, S11 increases from -2.34 GPa at the top surface to 647 MPa at a depth of 145 μm. This indicates that the observed residual stresses form a gradient in which the total mechanical response upon loading is driven by.

While in Fig. 17, the calculated S22 stress component for the AR sample remains close to zero, for the P6 and P9 samples, it strongly varies from the surface to the middepth. The values at the surface are 784 MPa and 760 MPa for P6 and P9, respectively. These features can develop only if the residual stress field introduced by LSP plays an important role in modifying the material stress state, particularly the surface stress state, which is of primary importance for understanding the phase transformation behavior of the treated NiTi alloy.

6.3. Effects of residual stresses on the strain distribution

With further evidence of residual effects due to the stresses, the strain distribution analysis results for the LE11 and LE22 components are shown in Figs. 18 and 19, respectively. The strains are quite uniform in the AR sample, with an LE11 of approximately 6.57% and an LE22 of approximately -2.17%. In contrast, for the P6 and P9 samples, the strain values in LE11 increase somewhat from the surface to midthickness, whereas they increase noticeably for LE22 from the surface to some depth before levelling off. This nonuniformity in strain distribution is most likely due to the different residual stress fields within the material, thus suggesting that residual stresses affect not only the phase transformation stress but also the whole deformation characteristic of the material.

6.4. Residual stress and transformation mechanism

Both the nonuniform stress/strain fields in the P6 and P9 samples result from the complex interaction between the residual stress and the phase transformation mechanism. The compressive residual stresses induced by LSP hinder dislocation movement and, consequently, martensite transformation nucleation in the surface region due to the reduced free energy difference for the driving force of transformation.

With further loading, however, reduced internal stress allows the phase transformation to occur at lower applied stresses, which assists the observed reduction in phase transformation stress.

Furthermore, the distributions of the von Mises stress (Fig. 15) and S11 stress (Fig. 16) suggest that the internal stress field should promote a more gradual phase transformation in the P6 and P9 samples than in the AR sample, in which the transformation is almost uniform within the cross section. This conclusion highlights the necessity of performing residual stress engineering to tune the mechanical behavior of NiTi SMAs in applications where the phase transformation stress needs to be strongly controlled.

6.5 Limitations and future work

While these numerical simulations are instructive and very enlightening, their limitations need to be put into perspective. The model assumes a simplified 2D plane stress condition that arguably cannot capture complexities of true 3D stress and strain distributions in real samples [20, 36]. Moreover, even though the calibrated material model fits the experimental data, it may not capture all the intricacies of martensitic transformation and reorientation under complex residual stress fields.

The work, therefore, will be channeled into 3D simulations to capture greater details of the distributions of stress and strain and their consequent effects on phase transformation. Experimental verification is also needed to a greater extent, particularly with respect to investigating different LSP treatment conditions and indexing the response to phase transformation behavior.

## 7. Conclusions

In this study, we explored the significant impact of residual stress on the isothermal tensile behavior of nanocrystalline NiTi-shaped memory alloys. Using a novel combination of focused ion beam (FIB) and digital image correlation (DIC), we successfully measured the two-dimensional residual stress in prestrain laser shock peened NiTi plates. The accuracy of this measurement technique was corroborated through a four-point bending experiment, reinforcing the reliability of the FIB-DIC materials. Our findings reveal that the internal residual stress plays a crucial role in reducing the phase transition stress of the NiTi SMA. This relationship was further elucidated through finite element analysis and theoretical analysis, offering a deeper understanding of the underlying mechanisms. These insights highlight the potential of residual stress engineering as a means to tailor the mechanical behavior of NiTi shape memory alloys, paving the way for optimized performance in practical applications. Overall, the integration of FIB-DIC for residual stress measurement and subsequent analysis represents a significant advancement in the field, providing a robust framework for future research and development in the mechanical behavior modulation of shape memory alloys.

**Acknowledgments**

The authors acknowledge the assistance of the SUSTech Core Research Facilities and Materials Characterization and Preparation Facility (MCPF) of HKUST.

**Declaration of conflicting interests**

The authors declare that there are no conflicts of interest.

**Funding**

This work was financially supported by the National Natural Science Foundation of China (Grant No. 12302095), Hong Kong Research Grant Council (RGC) through the GRF grant (Project No. 16212322), the Project of Hetao Shenzhen-Hong Kong Science and Technology Innovation Cooperation Zone (HZQB-KCZYB-2020083), the Science, Technology and Innovation Commission of Shenzhen Municipality of China (Project No. SGDX2019081623360564).